\documentclass{iopart}
\usepackage{mathptmx}
\usepackage{bm,amsfonts}
\makeatletter
\@ifundefined{pdfoutput} 
{ 
\usepackage[dvips]{graphicx,color}
}
{ 
\ifnum\pdfoutput=0\relax
\usepackage[dvips]{graphicx,color}
\fi
\ifnum\pdfoutput=1\relax
\usepackage[pdftex]{graphicx,color}
\fi
}
\makeatother
\newcommand{\beq}{\begin{equation}}
\newcommand{\eeq}{\end{equation}}
\newcommand{\beqa}{\begin{eqnarray}}
\newcommand{\eeqa}{\end{eqnarray}}
\newcommand{\nn}{\nonumber \\}
\def \e {\mathrm{e}}

\def \la {\langle}
\def \ra {\rangle}
\def \s {\sigma}

\def \B {{\mathcal B}}

\def \I {{\mathbb I}}

\def \M {{\mathcal M}}

\def \qh {\mathrm{qh}}

\begin{document}
\title[Computational equivalence]{Computational equivalence of the two inequivalent spinor representations of the 
braid group in the Ising topological quantum computer }
\author{Lachezar S. Georgiev}
\address{Institute for Nuclear Research and Nuclear Energy, Bulgarian Academy of Sciences,
72 Tsarigradsko Chaussee, 1784 Sofia, Bulgaria}
\address{Institut f\"ur Mathematische Physik, Technische Universit\"at Braunschweig,
Mendelssohnstr. 3, 38106 Braunschweig, Germany}
\begin{abstract}
We demonstrate that the two inequivalent spinor representations of the
braid group $\B_{2n+2}$,  describing the exchanges of $2n+2$ non-Abelian Ising anyons in the Pfaffian topological
quantum computer, are equivalent from computational point of view, i.e.,
the sets of topologically protected quantum gates that could be implemented in both cases by braiding
exactly coincide. 
We give the explicit matrices generating almost all braidings in the spinor representations of the $2n+2$ Ising anyons, 
as well as important recurrence relations.
Our detailed analysis allows us to understand better the physical difference
between the two inequivalent representations and to propose a process that could determine
the type of representation for any concrete physical realization of the
Pfaffian quantum computer. 
\end{abstract}
\section{Introduction}
Quantum computation \cite{nielsen-chuang,preskill-QC} has recently become an attractive
field of research because of the expected exponential speed-up over
the classical computations which could eventually allow us to perform hard computational
tasks that are practically impossible for classical computers.
One important class of systems in which quantum information is
believed to be protected from noise by topology is the class of topological
quantum computers
\cite{kitaev-TQC,freedman-kitaev-larsen-wang,preskill-TQC,sarma-freedman-nayak,TQC-PRL,TQC-NPB,sarma-RMP}.
One of the promising schemes for Topological Quantum Computation (TQC) is
based on the non-Abelian statistics of the low-energy quasiparticles in
low-dimensional strongly correlated systems such as the p-wave superconductor
\cite{read-green,ivanov,sarma-RMP} and the
Fractional Quantum Hall (FQH) state at filling factor $\nu=5/2$ in the second Landau level.
There is strong analytical and numerical evidence \cite{morf,rezayi-haldane} that the
$\nu=5/2$ FQH state belongs to the universality class of the Moore--Read (MR), or Pfaffian, 
 state \cite{greiter-wen-wilczek,mr,nayak-wilczek,CMP99,ivanov}.
The main idea of the TQC in the Pfaffian state
\cite{freedman-larsen-wang-braid,freedman-larsen-wang-TQC,sarma-freedman-nayak,Bravyi-5-2,TQC-PRL,TQC-NPB,sarma-RMP}
is to use the non-Abelian statistics of the Ising anyons to execute
quantum gates by adiabatically exchanging the quasiparticles comprising the Pfaffian qubits.
Because the Berry connection is trivial  \cite{stern-oppen-mariani,read-viscosity} the entire 
effect of the adiabatic transport comes from the explicit braiding and monodromy of the Ising conformal 
blocks.
The advantage of using FQH anyons for quantum computation is that the encoded
quantum information is naturally protected from
noise by topology because of the FQH energy gap which suppresses exponentially
all processes leading to noise and decoherence \cite{sarma-freedman-nayak}.

The elementary exchanges of adjacent quasiparticles in the Pfaffian FQH state
with $2n$ Ising quasiholes could be expressed \cite{nayak-wilczek} in terms of $\pi/2$-rotations from
the orthogonal group SO$(2n)$ and the corresponding braid operators belong to one of
the two inequivalent spinor representations $S_\pm$ of the covering group\cite{nayak-wilczek,read-JMP} 
Spin$(2n)$ of SO$(2n)$.
The fact that $S_\pm$ are inequivalent means that
 the two sets of matrices generating  $S_\pm$ differ by more than just a ``change of basis'',
which raises the following reasonable questions: 
if we assume that an experimental $\nu=5/2$ FQH  sample  is indeed in the universality class
of the MR state then in which of the two inequivalent spinor representations is this system
and how do we distinguish between them in a real FQH sample? 
Next, are these inequivalent representations  equivalent from
the computational point of view, i.e., could one construct the same number and types of
quantum gates in the two representations, or, the sets of quantum gates are significantly different?
In this paper we shall emphasize the physical difference and shall
proof that the two inequivalent representations of the braid group
$\B_{2n+2}$, are computationally equivalent.

It is worth stressing that as  a mathematical fact the computational equivalence of the two inequivalent 
Ising-type representations of the braid group $\B_{2n+2}$ is not a new result. It could be derived from the 
explicit representation of the monodromy subgroup in Ref.~\cite{franko}, corresponding to the universal 
R-matrix of the Ising model.
However, there is no proof in the literature that the Pfaffian wave functions with $2n+2$ Ising anyons
realize precisely  this representation of the braid group $\B_{2n+2}$ which is given in Ref.~\cite{franko}, 
though it is intuitively clear that it should be equivalent to that of the critical Ising model, yet this equivalence 
might be nontrivial and this is important for the physical implementation of various quantum gates. 
Therefore, in this paper we give an independent and self-consistent proof of the computational equivalence 
directly in the spinor approach of Ref.~\cite{nayak-wilczek}. 
A central result in this paper is  the derivation of a set of recurrence  relations for the braid 
generators $R_j^{(n,\pm)}$ of the spinor representations, which are necessary for the consistency of the proof, 
presented in Ref.~\cite{ultimate}, of the mathematical equivalence of the braid group representations derived 
from the Pfaffian wave functions and those obtained in the spinor approach.

The paper is organized as follows:
in Sect.~\ref{sec:clifford} we review the Clifford-algebra construction
\cite{nayak-wilczek} of the two inequivalent spinor representations of the
braid group $\B_{2n}$; in Sect.~\ref{sec:1qubit} we construct explicitly the
two inequivalent spinor representations of the braid group for 4 Ising anyons,
representing 1 qubit, and prove directly that they are computationally equivalent.
 In Sect.~\ref{sec:n-qubits}
we derive the recurrence relations and give explicit formulas for the braid generators in the 
positive- and negative-parity representations
of the  $n$-qubit systems in terms of $2n+2$ anyons, as well as 
a general proof of the computational equivalence of the two inequivalent
representations of the braid group $\B_{2n+2}$. Then,
in Sect.~\ref{sec:calibrate} we describe how to determine the type of a
concrete representation of the Pfaffian system with many non-Abelian anyons.
\section{Braid-group representation for Ising anyons in terms of the Clifford algebra }
\label{sec:clifford}
It was suggested in Ref.~\cite{nayak-wilczek} that the Ising-anyon representation of the braid group
$\B_{2n+2}$ can be realized by $\pi/2$ rotations from the group SO$(2n+2)$, however, this statement has not 
been completely proven in \cite{nayak-wilczek}. 
From the TQC point of view the most natural approach to compute the braid generators is to make analytic 
continuation of the $(2n+2)$-anyon CFT correlation functions which have been used originally  to construct 
the $n$-qubit Ising system \cite{nayak-wilczek,TQC-PRL,TQC-NPB}.
 However, in order to derive consistently all braid matrices for more Ising anyons one needs to know the Pfaffian 
wave functions in the negative-parity representation. This difficulty has been overcome in Ref.~\cite{ultimate}
where all braid generators in both representations with positive and negative parity have been consistently  derived 
by using the fusion rules in the Neveu--Schwarz and Ramond sectors of the Ising model. In addition it has been
finally proven\footnote{notice that this proof uses relations (\ref{i})--(\ref{iii}) given in Sect.~\ref{sec:n-qubits} below 
and without them it would be logically incomplete} in Ref.~\cite{ultimate} that the braid-group representation derived 
by analytic continuation of the multi-anyon Pfaffian  wave functions is indeed equivalent to one of the 
spinor representations of SO$(2n+2)$ and the equivalence matrices have been constructed explicitly for 
both representations with positive and negative parity. 

The elementary matrices representing the exchanges of $2n$ Ising quasiparticles in the
Pfaffian FQH state can be expressed \cite{nayak-wilczek} in terms of the gamma matrices
$\gamma_i^{(n)}$,  $1\leq i \leq 2n$, satisfying the anticommutation relations of the Clifford algebra
\beq \label{cr}
\left\{\gamma_i^{(n)},\gamma_j^{(n)} \right\} = 2\delta_{ij}, \quad 1\leq i,j \leq 2n.
\eeq
In more detail,
the elementary operations for the exchange of the $i$-th and $(i+1)$-th quasiparticles
could be expressed as \cite{nayak-wilczek,ivanov}
\beq \label{R}
R_{j}^{(n)} = \e^{i\frac{\pi}{4}}
\exp\left(-\frac{\pi}{4} \gamma_j^{(n)}\gamma_{j+1}^{(n)}\right) \equiv
\frac{\e^{i\frac{\pi}{4}}}{\sqrt{2}}
\left(\I -\gamma_j^{(n)}\gamma_{j+1}^{(n)}\right),
\eeq
where the second  equality follows from the fact that $(\gamma_j\gamma_{j+1})^2=-\I$
due to the anticommutation relations (\ref{cr}).

The $2n$  matrices $\gamma^{(n)}_i$ have dimension $2^n \times 2^n$ and
can be defined recursively in terms of the Pauli matrices
$\sigma_i$ ($i=1,2,3$) as follows  \cite{wilczek-zee}
\beqa \label{gamma}
\gamma_j^{(n+1)} &=&  \gamma_j^{(n)} \otimes \sigma_3 ,\quad 1\leq j \leq 2n \nn
\gamma_{2n+1}^{(n+1)} &=&  \I_{2^n}\otimes \sigma_1, \quad
\gamma_{2n+2}^{(n+1)} =  \I_{2^n} \otimes\sigma_2.
\eeqa
Starting with $n=0$ as a base, where $\gamma_{1}^{(1)} =\sigma_1$ and
$\gamma_{2}^{(1)} =\sigma_2$ we could write the gamma matrices explicitly as follows
\cite{wilczek-zee,nayak-wilczek,slingerland-bais}
\beqa \label{gamma-ex}
\gamma_{1}^{(n)} &=& \sigma_1 \otimes \sigma_3\otimes \cdots \otimes\sigma_3 ,  \quad
\gamma_{2}^{(n)} = \sigma_2 \otimes \sigma_3\otimes \cdots \otimes\sigma_3  \nn
&\vdots& \nn
\gamma_{2j-1}^{(n)} &=& \underbrace{\I_{2} \otimes \cdots \otimes \I_{2}}_{j-1}\otimes
\ \sigma_1 \otimes \underbrace{\sigma_3\otimes\cdots \otimes\sigma_3 }_{n-j}  \nn
\gamma_{2j}^{(n)} &=&  \underbrace{\I_{2} \otimes  \cdots \otimes \I_{2}}_{j-1}\otimes
\ \sigma_2 \otimes \underbrace{\sigma_3\otimes\cdots\otimes\sigma_3}_{n-j}  \nn
&\vdots& \nn
\gamma_{2n-1}^{(n)} &=&  \I_{2^{n-1}}\otimes \sigma_1 , \quad 
\gamma_{2n}^{(n)}  =  \I_{2^{n-1}} \otimes\sigma_2.
\eeqa
The ``gamma-five'' matrix $\gamma_F^{(n)}$,  defined by
$
\gamma_F^{(n)} =(-i)^n \gamma_1^{(n)} \cdots  \gamma_{2n}^{(n)},
$
commutes with all matrices (\ref{R}) and therefore the matrices $R_j^{(n)}$ cannot change the 
$\gamma_F^{(n)}$ eigenvalues $\pm 1$, which means that the representation (\ref{R}) is reducible
and the two irreducible components, corresponding to  eigenvalues $\pm 1$, can be obtained
by projecting with the two projectors 
\beq \label{P_pm}
P_{\pm}^{(n)} =\frac{\I_{2^n} \pm\gamma_F^{(n)} }{2}, \ \mathrm{i.e.,}
\ \left( P_{\pm}^{(n)}\right)^2= P_{\pm}^{(n)} =\left( P_{\pm}^{(n)}\right)^\dagger .
\eeq
In other words, the generators of the two irreducible spinor representations of the braid group
$\B_{2n}$ can be obtained by simply projecting\footnote{note that $P_{\pm}^{(n)}$ are even in the $\gamma$ 
matrices and therefore commute with $R_{j}^{(n)}$} Eq.~(\ref{R})
\beq \label{R_pm}
R_{j}^{(n,\pm) } =\e^{i\frac{\pi}{4}}
P_{\pm}^{(n)} \exp\left(-\frac{\pi}{4} \gamma_j\gamma_{j+1}\right)P_{\pm}^{(n)}   
= \frac{\e^{i\frac{\pi}{4}}}{\sqrt{2}}\left(\I -\gamma_j^{(n)}\gamma_{j+1}^{(n)}\right)
P_{\pm}^{(n)} . \,
\eeq
The elementary exchange matrices (\ref{R_pm}) are what we could eventually 
use to implement topologically protected quantum gates with Ising anyons.
\section{The single-qubit case: 4 Ising anyons }
\label{sec:1qubit}
In this section we shall explicitly  construct the two inequivalent spinor representations
of the braid group $\B_4$ following the general procedure described in Sect.~\ref{sec:clifford}.
The 4-dimensional $\gamma$ matrices in this case are explicitly
\beq \label{gamma_2}
\gamma^{(2)}_1 = \s_1\otimes \s_3 , \quad
\gamma^{(2)}_2= \s_2 \otimes \s_3 , \quad
\gamma^{(2)}_3 = \I_2 \otimes \s_1, \quad
\gamma^{(2)}_4= \I_2 \otimes \s_2 \quad
\eeq
and the diagonal matrix
$
\gamma_F^{(2)}=-\gamma^{(2)}_1\gamma^{(2)}_2\gamma^{(2)}_3\gamma^{(2)}_4
=\mathrm{diag}(1,-1,-1,1)
$
determines the two projectors on the two spinor representations $S_\pm$ to be
\beq \label{P2}
P^{(2)}_+=\mathrm{diag}(1,0,0,1) \quad \mathrm{and} \quad
P^{(2)}_-=\mathrm{diag}(0,1,1,0) .
\eeq
Applying the two projectors (\ref{P2}), and deleting the (zero) rows and columns
with numbers 2,3 for $P_+^{(2)}$ and 1,4 for $P_-^{(2)}$ respectively, we obtain
the three elementary generators of the 2-dimensional spinor representations $S_\pm$
of the braid group $\B_4$ as follows
\beq\label{R+}
R_{1}^{(2,+)}=\left[ \matrix{ 1 & 0 \cr 0 & i}\right],
R_{2}^{(2,+)}=\frac{\e^{i\frac{\pi}{4}}}{\sqrt{2}}
\left[ \matrix{ 1 & i \cr i & 1}\right],
R_{3}^{(2,+)}=\left[ \matrix{ 1 & 0 \cr 0 & i }\right],
\eeq
and
\beq \label{R-}
R_{1}^{(2,-)}=\left[ \matrix{ 1 & 0 \cr 0 & i }\right],
R_{2}^{(2,-)}=\frac{\e^{i\frac{\pi}{4}}}{\sqrt{2}}
\left[ \matrix{ 1 & \!\!-i \cr \!\!-i & 1}\right],
R_{3}^{(2,-)}=\left[ \matrix{ i & 0 \cr 0 & 1}\right].
\eeq
\textbf{Remark 1.} \textit{The positive-parity representation of $\B_4$,  generated by $R_j^{(2,+)}$,
looks different from that obtained in Ref.~\cite{ultimate} by analytic continuation of the 4-anyon 
Pfaffian wave functions (with generators $B_j^{(4,+)}$ there), however, as proven in \cite{ultimate}, these 
two positive-parity representations are equivalent and the matrix establishing 
this equivalence is simply $Z=\mathrm{diag}(1,-1)$.} \\

It is worth-stressing that the two inequivalent representations $S_\pm$ of the braid group
$\B_4$, generated from the elementary braid matrices  (\ref{R+}) and (\ref{R-}) respectively
and their inverses coincide as sets of matrices. This is because as we saw before
$R_{1}^{(2,+)}=R_{1}^{(2,-)}$, and because
\[
R_{2}^{(2,+)} R_{2}^{(2,-)}=R_{3}^{(2,+)} R_{3}^{(2,-)}= i \I_2 .
\]
Note that the matrix $i\I_2$ does belong to both representations of the braid group $\B_4$, i.e.,
\beq
R_{1}^{(2,\pm)}R_{2}^{(2,\pm)}\left(R_{3}^{(2,\pm)}\right)^2
R_{2}^{(2,\pm)}R_{1}^{(2,\pm)}=\pm i\I_2 .
\eeq
In other words all 2-dimensional matrices which can be obtained by braiding Ising anyons
in the representation $S_+$ can be implemented in the representation $S_-$  as well
so that the two inequivalent representations $S_\pm$ are computationally equivalent.

It is not difficult to see that the diagonal elementary braid matrices in each of the representations 
$S_\pm$ of  $\B_4$  fix the single-qubit computational basis (up to equivalence). Indeed, consider the matrix
$R_{1}^{(2,+)}$: before braiding we can first fuse the quasiholes at positions $\eta_1$ and
$\eta_2$ by using the fusion rule \cite{fst,TQC-NPB}
\beq \label{fusion}
\psi_\qh(\eta_1) \psi_\qh(\eta_2) \mathop{\simeq}\limits_{\eta_1\to \eta_2}
\left(\I+\frac{1}{\sqrt{2}}\sqrt{\eta_{12}} \ \psi(\eta_2) \right)
\e^{i\frac{\phi(\eta_2)}{\sqrt{2}}},
\eeq
where $\I$ corresponds to the fusion channel $\s_+\s_+\simeq |0\ra$  while
the Majorana fermion $\psi$ corresponds to the fusion channel $\s_+\s_-\simeq |1\ra$.
Here $\s_\pm$ are the chiral spin fields of CFT dimensions $1/16$ of the Ising model
\cite{milo-read,fst,5-2,TQC-PRL,TQC-NPB} and the subscript $\pm$ denotes their fermion parity.
Executing the braid is now equivalent to the transformation
$\eta_{12} \to \e^{i\pi}\eta_{12}$  so that the resulting phase is 1 if the
first pair is $\s_+(\eta_{1})\s_+(\eta_{2})$ and $i$ if it is
$\s_+(\eta_{1})\s_-(\eta_{2})$
 and this topological phase is independent
of how close are the two anyons.
Therefore, the braid matrix $R_{1}^{(2,+)}$ completely  determines the
type of the $\s$ fields with coordinates $\eta_1$ and $\eta_2$.
In the same way if we braid $\eta_3$ with $\eta_4$, i.e.,
$\eta_{34} \to \e^{i\pi}\eta_{34}$ the pair $\s(\eta_3) \s(\eta_4)$ obtains a phase 1
if it is in the channel of the identity or $i$  if it is in the channel of $\psi$.
Thus, from the explicit form of the diagonal matrices $R_{1}^{(2,+)}$ and $R_{3}^{(2,+)}$,
 we can unambiguously reconstruct the single-qubit computational basis in terms of the Ising-model 
correlation functions for the spinor representation $S_+$ as follows
\beq \label{1qb-basis1}
|0\ra_+ \equiv \la \s_+\s_+ \ \s_+\s_+   \prod_{j=1}^{2N} \psi(z_j) \ra, \quad
|1\ra_+ \equiv \la \s_+\s_- \ \s_+\s_-  \prod_{j=1}^{2N} \psi(z_j) \ra  .
\eeq
Similarly, from the diagonal matrices $R_{1}^{(2,-)}$ and
$R_{3}^{(2,-)}$ in Eq.~(\ref{R-}),  we can unambiguously reconstruct the
single-qubit basis in the spinor representation  $S_-$
\beq\label{1qb-basis2}
|0\ra_- \equiv \la \s_+\s_+ \ \s_+\s_-   \prod_{j=1}^{2N+1} \psi(z_j) \ra, \quad
|1\ra_- \equiv \la \s_+\s_- \ \s_+\s_+   \prod_{j=1}^{2N+1} \psi(z_j)  \ra  .
\eeq
The above analysis clarifies the physical difference between the two inequivalent
spinor representations: the representation $S_+$ is realized (in the large-$N$ limit) with 4 $\s$ fields, with 
positive total parity and even number of Majorana fermions, while $S_-$ corresponds to 4 $\s$ fields, with 
negative total parity and odd number of Majorana fermions.
\section{The $n$-qubit case: projected braid matrices for $2n+2$ Ising anyons}
\label{sec:n-qubits}
Using the recursive definition (\ref{gamma}) of the gamma matrices
one can directly prove that most of the unprojected exchange matrices
 for $2n+2$ anyons  are simply related to those for  $2n$ anyons
\beq \label{most}
	R_{j}^{(n+1)}=  R_{j}^{(n)} \otimes \I_2\quad \mathrm{for} \quad
	1\leq j \leq 2n-1,
\eeq
where the superscript $(n)$ or $(n+1)$ now refers to the superscript of the corresponding
gamma matrices entering Eq.~(\ref{R}).

Next, because of the recursive relation 
$
\gamma_F^{(n+1)} =  \gamma_F^{(n)} \oplus \left(-\gamma_F^{(n)}\right),
$
where $\oplus$ denotes the direct sum of matrices,
it is easy to prove that the projectors (\ref{P_pm}) are also recursively related by
\beq \label{P-recurrence}
P^{(n+1)}_+ = P^{(n)}_+ \oplus P^{(n)}_-, \quad
P^{(n+1)}_- = P^{(n)}_- \oplus P^{(n)}_+ .
\eeq
\subsection{Recurrence relations for the projected braid matrices}
\label{sec:recurrence}
We can now prove that the projected matrices (\ref{R_pm}) satisfy the following recurrence relations
\beqa \label{i}
R_j^{(n+1,+)}&=&R_j^{(n+1,-)} \qquad \textrm{for} \quad  1\leq j \leq 2n-1 \\
\label{ii}
R_j^{(n+1,\pm)}&=&R_j^{(n,\pm)} \otimes \I_2 \qquad \textrm{for} \quad 1\leq j \leq 2n-3
\\
\label{iii}
R_j^{(n+1,\pm)}&=&R_{j-2}^{(n,\pm)} \oplus R_{j-2}^{(n,\mp)} \quad \textrm{for} \quad 3\leq j \leq 2n+1, \quad
\eeqa
which together with the two-qubit case, $n=2$, as a base\footnote{the validity of the recurrence 
relations (\ref{i})--(\ref{iii})  for $n=2$ could be directly checked from Eq. (\ref{R_pm})}  completely 
determine all projected matrices (\ref{R_pm}).
To prove Eq.~(\ref{ii}) notice that due to the structure of the projectors (\ref{P-recurrence}) we have
\[
P^{(n+1)}_{\pm}=\mathop{\oplus}\limits_{i=1}^{2^n} P^{(1)}_{\pm \alpha(i)},\quad \mathrm{where}
\quad \alpha(i)=\pm
\]
so that $P_\pm^{(n+1)}$ can be written as block-diagonal matrices whose elements on the diagonal are
the $2\times 2$ dimensional matrices
\beq \label{P_1}
P^{(1)}_{+} = \left[ \begin{array}{cc} 1 & 0 \cr
0 & 0 \end{array}\right] \quad \mathrm{and} \quad
P^{(1)}_{-} = \left[ \begin{array}{cc} 0 & 0 \cr
0 & 1 \end{array}\right].
\eeq
Applying any of these projectors to the unprojected matrices (\ref{most}) simply removes the 
matrix $\I_2$ from the tensor product in Eq.~(\ref{most}), i.e.,
\[
R_{j}^{(n+1,+)}=  R_{j}^{(n+1,-)} = R_{j}^{(n)} \quad \textrm{for}  \quad 1\leq j \leq 2n-1,
\] 
which proves Eq.~(\ref{i})  and expresses the trivial relation between 
projected matrices for $n$ qubits to the unprojected matrices for $(n-1)$ qubits for this values of $j$. 
If there is one more $\I_2$ in 
the unprojected matrix (\ref{most}), as in the case when  $1\leq j \leq 2n-3$, this relation proves Eq.~(\ref{ii}).
On the other hand, when $3\leq j \leq 2n+1$ the unprojected matrices are again tensor products in which, however, 
the unit matrix is to the left
\[
R_{j}^{(n+1)}=  \I_2\otimes R_{j-2}^{(n)}  =  R_{j-2}^{(n)} \oplus  R_{j-2}^{(n)}  ,
\quad 3\leq j \leq 2n+1,
\]
so that applying the projectors (\ref{P-recurrence}) proves Eq.~(\ref{iii}).
Notice the index shift $j\to j-2$ which is due to the relabeling $\eta_j'=\eta_{j-2}$, for $3\leq j \leq 2n+1$,
after splitting one unit matrix $\I_2$ to the left corresponding to the first qubit encoded into the pair of anyons 
with coordinates $\eta_1$ and $\eta_2$.\\

\noindent
\textbf{Remark 3.} \textit{
The recurrence relations (\ref{i})--(\ref{iii})  for $R_j^{(n+1,\pm)}$ are identical with relations (28)--(30) in 
Ref.~\cite{ultimate}, for $B_j^{(2n+2,\pm)}$ (see \cite{ultimate} for the notation) 
despite the fact that $R_j^{(n+1,\pm)}$ and $B_j^{(2n+2,\pm)}$
generate in principle different representations of the braid group $\B_{2n+2}$, 
and this fact has been used in the proof of the Nayak--Wilczek conjecture there.
Without  Eqs.~(\ref{i})--(\ref{iii}) the proof of the Nayak--Wilczek conjecture in Ref.~\cite{ultimate} would be 
logically incomplete.}
\subsection{Explicit formulas for the projected braid matrices}
\label{sec:explicit}
Using Eq.~(\ref{gamma-ex}) for the  $\gamma$ matrices, and relations (\ref{i})--(\ref{iii}), we can 
obtain most of the projected  braid matrices $R_j^{(n+1,\pm)}$ explicitly 
\beq \label{S_j}
R_{2j-1}^{(n+1,\pm)}= \underbrace{\I_2\otimes \cdots \otimes \I_2}_{j-1} \otimes 
\left[ \begin{array}{cc}1 & 0 \cr 0 & i\end{array}\right] \otimes \underbrace{\I_2\otimes \cdots \otimes \I_2}_{n-j} ,
\eeq
for $1\leq j\leq n$,
\beq \label{R_2j}
R_{2j}^{(n+1,\pm)}= \underbrace{\I_2\otimes \cdots \otimes \I_2}_{j-1} \otimes 
\frac{\e^{i\frac{\pi}{4}}}{\sqrt{2}}
\left[ 
\begin{array}{cccc}1 & 0 & 0 & i \cr 0 & 1 & -i & 0 \cr 0 & -i & 1 & 0 \cr i & 0 & 0 & 1 \end{array}\right]  
\otimes  \underbrace{\I_2\otimes \cdots \otimes \I_2}_{n-j-1} ,
\eeq
for $n\geq 2$ and  $1\leq j\leq n-1$.
These has to be supplemented by the recurrence relations for the last two generators
which do not have a tensor product structure, however, still can be constructed recursively from
\beq \label{R_2n}
R_{2n}^{(n+1,\pm)}= R_{2n-2}^{(n,\pm)}\oplus R_{2n-2}^{(n,\mp)}, \quad 
R_{2n+1}^{(n+1,\pm)}=R_{2n-1}^{(n,\pm)}\oplus R_{2n-1}^{(n,\mp)}
\eeq
using as a base  the matrices $R_2^{(2,\pm)}$ and $R_{3}^{(2,\pm)}$ defined in Eqs.~(\ref{R+}) and (\ref{R-}).
 Equations (\ref{S_j}),  (\ref{R_2j}) and (\ref{R_2n})  give the most explicit and practical
description of  the projected braid generators $R_j^{(n+1,\pm)}$.
\subsection{Proof of the computational equivalence}
\label{sec:proof}
It is obvious from Eqs. (\ref{S_j}) and (\ref{R_2j}) that  $R_{j}^{(n+1,+)}=R_{j}^{(n+1,-)} $ for $1\leq j \leq 2n-1$,
as stated in Eq.~(\ref{i}), so that the two inequivalent spinor representations of $\B_{2n+2}$ 
differ only in the last two generators, i.e.,  for $j=2n, \, 2n+1$. 
It can now  be proven that the last two braid generators $R_{2n}^{(n+1,\pm)}$ and $R_{2n+1}^{(n+1,\pm)}$ 
in the two representations with opposite parity are mutually inverse up to an overall factor of $i$, i.e., 
\beq \label{last-two}
R_{2n}^{(n+1,+)}R_{2n}^{(n+1,-)} = i \I_{2^n}
\quad \mathrm{and} \quad 
R_{2n+1}^{(n+1,+)} \ R_{2n+1}^{(n+1,-)} =i \I_{2^n} .
\eeq
Indeed, it follows from Eqs. (31) and (32) in Ref.~\cite{clifford} that   
$\left(R_{j}^{(n,+)}\right)^2=-\left(R_{j}^{(n,-)}\right)^2 $ for
$j=2n, \, 2n+1$  (while for the other values of $j$ the squares of the braid generators in the two representations 
coincide).
Next, combining Eq. (35) in Ref.~\cite{clifford} with Eq.~(\ref{R_pm}) in this paper, it is easy to prove 
that (cf. Remark 2.5 in Ref.~\cite{franko}), 
\[
R_j^{(n+1,\pm)} = \frac{\e^{i\frac{\pi}{4}}}{\sqrt{2}}\left(\I_{2^n} - i \left(R_{j}^{(n+1,\pm)}\right)^2 \right),
\quad 1\leq j \leq 2n+1.
\]
Applying this equation for $j=2n, \, 2n+1$, and using the above results for the squares of the last two 
braid generators, we have
\beqa
R_{j}^{(n+1,+)}R_{j}^{(n+1,-)} &=&  \frac{i}{2} \left(\I_{2^n} - i \left(R_{j}^{(n+1,+)}\right)^2 \right)
\left(\I_{2^n} + i \left(R_{j}^{(n+1,+)}\right)^2 \right) \nn
 &=& \frac{i}{2} \left(\I_{2^n} - i^2 \left(R_{j}^{(n+1,+)}\right)^4 \right) =i \I_{2^n}  , \nonumber
\eeqa
for $j=2n, \, 2n+1$.
The last equality in the above equation follows from the fact that the unprojected generator could be 
interpreted as a rotation on $\pi/2$ so that $(R_{j}^{(n+1)})^4 =\I_{2^{n+1}}$,  hence 
the fourth power of the projected generators is 
$(R_{j}^{(n+1,\pm)})^4 =\I_{2^{n}}$ (cf. Eq.~(35) in \cite{clifford}).

The computational equivalence between the two inequivalent spinor representations $S_\pm$
of $\B_{2n+2}$ formally follows from Eqs.~(\ref{i}) and (\ref{last-two}) because the element $i\I_{2^n}$ 
is always an element of the monodromy group
$\M_{2n+2} \subset \B_{2n+2}$ in the Ising-model representation (see Eq.~(33) in Ref.~\cite{clifford}). 

\begin{figure}[htb]
\centering
\includegraphics*[bb=35 595 390 685,width=8.5cm]{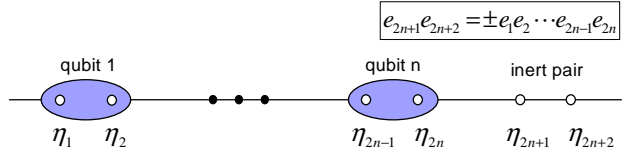}
\caption{$n$-qubit configurations in terms of $2n+2$ Ising quasiholes
	 corresponding  to the two spinor representations
of the braid group $\B_{2n+2}$. The sign $\pm$ in the charge conservation equation in the box
corresponds to the parity of the representation $S_\pm$. }
\label{fig:n-qubits}
\end{figure}

Again the explicit construction of the projected braid matrices for $n$ qubits, from the generators of
the spinor representations of SO$(2n+2)$, assumes a particular basis of computational states.
It is not difficult to see that,  like in the one-qubit case considered in Sect.~\ref{sec:1qubit},
the projected diagonal braid matrices (\ref{S_j}) and (\ref{R_2n}) completely fix the
$n$-qubit's computational bases in the two inequivalent representations.
As obvious from Eq.~(\ref{S_j}), the diagonal matrices $R_{2j-1}^{(n+1,\pm)}$
with  $1\leq j \leq n$, represents the phase gate $S=\mathrm{diag}(1,i)$
on the $j$-th qubit ($1\leq j \leq n$), so that the state of the $j$-th qubit
corresponding to the $\gamma$-matrices realization in Eq.~(\ref{gamma}),
 is determined by the pair of
Ising anyons  with  coordinates $\eta_{2j-1}$ and $\eta_{2j}$, and  the qubits are
ordered from left to right as shown in Fig.~\ref{fig:n-qubits}.
The last two anyons, with the coordinates $\eta_{2n+1}$ and $\eta_{2n+2}$,
form the inert pair which is responsible for restoring the total fermion parity in the
two inequivalent representations.
In other words, the explicit definition of the SO$(2n+2)$ $\gamma$ matrices, as in Eqs.~(\ref{gamma}) and 
(\ref{gamma-ex}), already assumes the structure and the ordering of the $n$-qubit system as in Fig.~\ref{fig:n-qubits}.
\section{Calibration of the Pfaffian quantum computer}
\label{sec:calibrate}
\begin{figure}[htb]
\centering
\includegraphics*[bb=65 620 500 780,width=8.5cm]{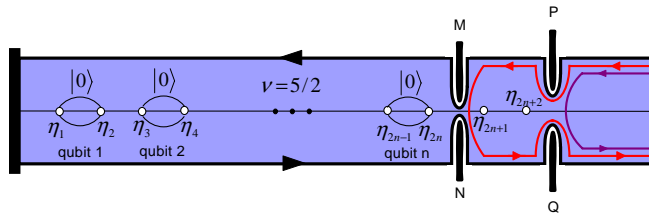}
\caption{Calibrating the Ising-anyon quantum computer by measuring the topological charge of the last pair of Ising
anyons.
}
\label{fig:calibrate}
\end{figure}

The analysis performed here allows us to unambiguously fix the type of the spinor
representation in a real physical sample, calibrating in this way the
Pfaffian quantum computer. To this end we propose the following procedure
for $2n+2$ Ising anyons corresponding to $n$ qubits:

\begin{enumerate}
\item Initialize the $n$-qubit system in the state
$|00\cdots 0\ra$. This could be done by applying the single-qubit initialization
scheme of Das Sarma et al. \cite{sarma-freedman-nayak} for each pair of anyons.
\item Measure the total topological charge of the system  by Fabry-Perot interferometry
\cite{fradkin-nayak-tsvelik-wilczek,Fabry-Perot-exp,sarma-freedman-nayak}. 
This charge would be $+1$ if the system is in the 
representation $S_+$ and   $-1$ if it is in $S_-$.
Because all qubits are in the state $|0\ra$ the total topological charge is equal
to the topological charge of the last pair of Ising quasiholes with coordinates
$\eta_{2n+1}$ and $\eta_{2n+2}$.
We can therefore determine the topological charge of the pair $(\eta_{2n+1},\eta_{2n+2})$
as shown in Fig.~\ref{fig:calibrate} like it was done in the original approach of
Ref.~\cite{sarma-freedman-nayak}.
\end{enumerate}
\ack
I thank Chetan Nayak, Lyudmil Hadjiivanov, Sergey Bravyi, Shankar Das Sarma and 
Andr\'e Ahlbrecht 
for useful discussions. I also thank the organizers of the Statphys 23 Satellite Meeting in
Perugia 2007 for hospitality and financial support.
The author has been supported as a Research Fellow by the Alexander von Humboldt
foundation. This work has been also partially supported by the BG-NCSR under
Contract No. DO-02-257.
\bibliography{Z_k,my,FQHE,TQC}
\end{document}